\documentclass[12pt,thmsa]{article}
\usepackage{sw20lart}
\usepackage{amsmath}
\usepackage{graphicx}
\usepackage{amsfonts}
\usepackage{amssymb}

\begin{document}

\title{\vspace*{-0.5cm}Surface superconducting states and paramagnetism in mesoscopic
superconductors }
\author{C. Meyers\thanks{email: meyers@cpmoh.u-bordeaux.fr}\\Condensed Matter Theory Group{\small , }CPMOH\\CNRS UMR-5798, Universit\'{e} Bordeaux 1, \\33405 Talence Cedex{\small , }France}
\date{}
\maketitle
\begin{abstract}
In the framework of the Ginzburg-Landau equation, the temperature dependence
of the upper critical field of small ring-like superconductors is studied. At
equilibrium small parts of the phase diagram show paramagnetism for width /
radius ratios below 0.85. Their number and extension increase with the size of
the hole. In these regions, only the inner part of the ring shows a positive
magnetic moment. The order parameter density profile appears to change, when
crossing a first order transition line, which separates different angular
momentum values, and we clarify the relationship between the localization of
superconductivity nucleation and paramagnetism of those samples.
\end{abstract}

\section{ Introduction}

\bigskip

The paramagnetic Meissner effect (PME) observed in small superconducting
samples simply expresses the fact that the energy \ of those samples decreases
upon an increasing magnetic field. \ A reentrant (H,T) phase diagram is one of
the signals that characterizes this effect. This surprising effect has been
seen both in conventional\cite{ExpPara}\cite{Geim} and in high Tc
superconductors \cite{ExpParaHTc}.

The behavior of nanometer-scale superconducting grains\cite{Ralph}, which
shows Pauli paramagnetism\cite{GrainsP} in addition to other striking
properties of finite electron systems, is outside the scope of this work which
relies on the Ginzburg-Landau equation.

Metastability is one of the major ingredients that have been put forward to
explain PME in conventional superconductors \cite{metapapers}. For example, in
the Ginzburg-Landau approach of small size cylindrical samples at the upper
critical field, the disbalance of the screening currents leads to a positive
magnetic moment \cite{Zarkov}. But, in this case if the equilibrium
transitions are only allowed, the PME disapears \cite{Zarkov}, and many
mechanisms causing non equilibrium flux configuration have been proposed
\cite{metapapers}.

However PME is known to occur for minimal energy configurations of some
mesoscopic objects:\ the Little-Parks \cite{LitParks} ring is the most simple
example that shows an alternating sequence of diamagnetic and paramagnetic
responses as the magnetic field is increased. On the theoretical side this
general behavior is known to be shared by all loop-like samples and moreover
that PME appears the be stable against thermic fluctuations \cite{DauMeyBuz}.

In the framework of the London approximation of the vortex phase of small
superconducting disks, we have obtained magnetization curves which showed a
PME, in good agreeement with the experimental observation \cite{Geim}, and
made the analogy with the Little-Parks ring to interpret such a behavior of
stable configurations \cite{MeyDau}.

The aim of this paper is to give the reader a simple example of a mesoscopic
ring-like superconducting film of which the paramagnetic response of the
stable configurations can be tuned with the size of the central hole, and to
show at which part of the ring (inner or outer) the nucleation of
superconductivity is localized.

In Section 2 the upper critical field $H_{c3}^{\ast}$ is calculated in the
framework of the Ginzburg-Landau equation, and, if the size of the hole is
large enough, the parts of the phase diagram where PME takes place are found.
In Section 3 we show the two regions of the ring with opposite supercurrents
which might give rise to a global positive magnetic moment. In Section 4 we
give in more details our results for a ring with a small hole with
$width\,/\,Radius=0.75$ for which there is only one paramagnetic domain and we
also show the order parameter density profile that changes drastically when
crossing the first order line. Our conclusions are drawn in the last section.

\section{Nucleation of superconductivity near the edges of a ring}

\bigskip

We shall follow quite closely the\ method of Bezryadin, Buzdin and Pannetier
\cite{BBP} \ in the case of a circular hole or of a disk\cite{BGRW}, in order
to obtain the phase diagram of a circular ring of external radius $R$ and
width $d$ under a uniform magnetic field $H$ normal to the plane of the film sample.

\bigskip In the case of \ a thin film of thickness $\epsilon$ the effective
screening length, $\lambda_{eff}=\lambda^{2}/\epsilon$ , becomes very large
and we may neglect the magnetic field energy in the Ginzburg-Landau (GL) free
energy functional:%

\[
F=%
{\displaystyle\int}
\left(  a\left|  \Psi\right|  ^{2}+\frac{b}{2}\left|  \Psi\right|  ^{4}%
+\frac{1}{4m}\left|  \left(  -i\hslash\overrightarrow{\mathbf{\nabla}}%
-\frac{2e}{c}\overrightarrow{\mathbf{A}}\right)  \Psi\right|  ^{2}\right)  dV
\]
\ \ 

where $a=\alpha(T-T_{0})$.

Moreover due to the small size of the sample, we can take for the vector
potential $\overrightarrow{A}$ the potential of the uniform external applied
field $H$ :

\begin{center}%
\[
\overrightarrow{\mathbf{A}}=\frac{1}{2}(\overrightarrow{\mathbf{H}}%
\times\overrightarrow{\mathbf{r}}\mathbf{)}
\]
\end{center}

Using polar coordinates ($\rho$ ,$\,\,\varphi$), the equation for the
normalized order parameter $\psi$ \ is:%

\[
\frac{d^{2}}{dx^{2}}\psi+\frac{1}{x}\frac{d}{dx}\psi-\left[  \frac{i}{x}%
\frac{\partial}{\partial\varphi}+\frac{\phi}{\phi_{0}}x\right]  ^{2}\psi
+\frac{R^{2}}{\xi^{2}}(\psi-\psi^{3})=0
\]

\bigskip where $\ x=\rho/R$ $\ \ \ $, $\ \ \psi=\Psi\,/\sqrt{\left|  a\right|
\,/\,b}$ , \ $\xi^{2}=\hslash^{2}/\,4\,m\,\left|  a\right|  $, $\phi_{0}%
=\pi\hslash\,c\,/\,e$, and $\,\phi=\pi\,R^{2}\,H$ \ is the flux of the
magnetic field across the plain disk (with no hole).

In order to obtain the upper critical field $H_{c3}^{\ast}$ \cite{SJdG}, using
the linearized version of the GL equation, we search for solutions of definite
angular momentum $n$ of the form:%

\[
\psi_{n}(x,\varphi)=f_{n}(x)\,\exp(in\varphi).
\]

\bigskip

Then, in the present gauge, the super-current is tangent to the two circular
insulating boundaries of the ring, and we may take the boundary conditions of
the GL equation as follows:%

\[
\frac{d}{dx}f{\LARGE |}_{x=1}=0\text{ , and \ \ \ }\frac{d}{dx}f{\LARGE |}%
_{x=1-d/R}=0\text{ .}
\]

After reduction to Kummers's equation, the general solution can be expressed
as the linear combination of two confluent hypergeometric functions $M$ and
$U$ \cite{Abra} as follows:

\begin{center}%
\[
f_{n}(x)=x^{n}\exp(-\frac{x^{2}}{2}\frac{\phi}{\phi_{0}}){\LARGE \bigskip
\ }\left\{  A\,\,M(Y,n+1,x^{2}\frac{\phi}{\phi_{0}})+B\,\,U(Y,n+1,x^{2}%
\frac{\phi}{\phi_{0}})\right\}
\]
{\LARGE \ \ }
\end{center}

where \ $Y=\frac{1}{2}-\frac{1}{4}\frac{R^{2}}{\xi^{2\,}}\frac{\,\phi_{0}%
}{\,\phi}$ , and where the two constants $A$ and $B$ have to be determined
using the boundary conditions.

In practice, we have first obtained the ratio $A/B$ in terms of $\phi$ ,
$R^{2}\,/\,\xi^{2}$ and $n$ using the first condition, while the second
condition has been numerically solved and allows us to obtain $R^{2}%
\,/\,\xi^{2}$ versus $\phi$ \ for each $n$. Then, for a given flux $\phi$ ,
the critical line $T_{c}(\phi)$ is obtained by choosing the angular momentum
value which minimizes \ $R^{2}\,/\,\xi^{2}$ \cite{BBP}. The limiting cases of
either a plain disk or a circular hole can be reached, using the same
procedure, with $B=0$ in the first case and with $A=0$ in the second one. The
Little-Parks ring \cite{LitParks} , can be recovered using a very small value
of $d/R$ , as shown on figure 1 . We see that paramagnetism starts just after
the transition between angular momentum states $J-1\rightarrow J$ and
disapears for $\phi/\phi_{0}\geq J$, until the next transition occurs. On
figure 2 we give the $H_{c3}^{\ast}$ line for a ring with $d\,/\,R=0.5$. The
cusps, associated to jumps of the angular momentum, are clearly seen and three
small paramagnetic domains are located above these transitions. For sake of
comparison the full disk $H_{c3}^{\ast}$ curve is shown on figure 3 , on which
no such domains are seen; such oscillations have been observed on micron-sized
superconducting disks\cite{BGRW}. Our results give $d\,/\,R\,\ \thickapprox
\ \ 0.85$ as the value under which paramagnetic domains are observed.

\bigskip

\section{Magnetic moment and supercurrents}

\smallskip In the state $f_{n}$ of angular momentum $n$, the magnetic moment
$M=-\partial F\,/\,\partial H$ \ can be written as: \ %

\[
M=\frac{\phi_{0}}{(2\pi\lambda)^{2}}\pi\,R^{2}\,\epsilon_{film}\,\int
_{1-d/R}^{1}\,f_{n}(x)^{2}\,\,\,(n-\frac{\phi}{\phi_{0}}\,x^{2})\,x\,dx.
\]

\thinspace In fact the local density of magnetic moment is related to the
value of the supercurrent:
\[
\mathbf{J}=\frac{\hslash}{4m}(\Psi^{\ast}\mathbf{\nabla}\Psi-\Psi
\mathbf{\nabla}\Psi^{\ast})-\frac{e}{2mc}\mathbf{A}\left|  \Psi\right|  ^{2}
\]

of which the components are:%

\[
J_{\varphi}=\frac{f_{n}(x)^{2}}{x}\,\,\,(n-\frac{\phi}{\phi_{0}}%
\,x^{2})\,\,\text{\ \ \ and \ \ \ }J_{\rho}=0
\]

For a given flux and angular momentum, and when the ring is crossed from the
inner edge ($x=1-d/R$) up to the outer one ($x=1)$, three various situations
can be encountered acording to the sign of $J_{\varphi\text{ \ }}$:$\newline $

\begin{itemize}
\item \bigskip The radial supercurrent density is negative everywhere on the
ring for $\frac{\phi}{\phi_{0}}>\frac{n}{(1-d/R)^{2}}$ . This gives a uniform
standard diamagnetic response.

\item \bigskip The radial supercurrent density is positive everywhere on the
ring for $\frac{\phi}{\phi_{0}}<n$ \ \ and gives a uniform paramagnetic response

\item  The current flows changes its sign at a point \ $x_{0}\,=\sqrt
{\frac{\phi_{0}}{\phi}\,n}$ \ \ . Thus one can distinguish two different
regions on the ring: the inner part which contibutes positively to the total
magnetic moment and the outer part which gives a diamagnetic contribution,
with currents in the two regions flowing in opposite directions.
\end{itemize}

\bigskip

Let us comment Figures 1-3 on this point. The ideal Little-Parks ring of Fig1
(with $d/R\thickapprox0$) shows a sequence of uniform responses: $J=0$ is
diamagnetic, whereas the reentrant part of $J=1$ is paramagnetic, etc... and
the supercurrent changes its sign for all half-integer values of $\frac{\phi
}{\phi_{0}}$ . Fig2, which depicts the behavior of a thicker ring (with
$d/R=0.5$), shows that apart the standard uniform diamagnetic response of the
$J=0$ state, the $J=1$ state gives a \ uniform paramagnetic response for small
fields $\frac{\phi}{\phi_{0}}<1$ \ . The magnetization of all the other states
results of the competition between the inner and the outer currents. The case
of the plain disk (Fig3) is interesting although no paragnetism is observed on
the global magnetization: all the non-zero angular momentum states have a
central paramagnetic response and a larger diamagnetic one located on the
periphery of the disk.

\bigskip

\section{\bigskip Paramagnetism in a disk with a small hole}

\bigskip

In this section, we have choosen to treat more extensively the particular case
of a small hole, $d\,/\,R=0.75$ , by solving the non-linear Ginzburg-Landau
equation. This allows us to enter the superconducting region, and compute the
densities of both the order parameter and the magnetization . In fact it can
be shown that mixture of different angular momentum states can be neglected if
only the lowest energy is searched for, so that we shall not consider such mixtures.

Because the two boundary conditions restrict $\psi$ at two different points,
namely $x=1-d/R$ and $x=1$, we have solved the non-linear GL equation \ in two
steps. We have used standard numerical routines which allowed us to obtain
$f_{n}(x)$ given the initial values $f_{n}(1)$ and $df_{n}\,/dx|_{x=1}=0$ ,
then $f_{n}(1)$ is found by solving $df_{n}\,/dx|_{x=1-d/R}=0$ \ for
$f_{n}(1)$.

The $H_{c3}^{\ast}$ line for the ring with $d\,/\,R=0.75$ is drawn on Fig. 4.
Only one small paramagnetic domain shows up for fluxes $1.4\lesssim\phi
/\phi_{0}\,\lesssim\,1.8$ with $J=1$. On the inset of Fig 4, which is a zoom
of the paramagnetic region, we have drawn a fixed temperature path (A-B) with
$R^{2}/\xi^{2}=1.2$ which crosses the first order transition line separating
the $J=0$ and $J=1$ phases. A fixed flux path (C-B) with $\phi/\phi
_{0}\,=1.625$ which lies entirely in the paramagnetic phase is also shown. The
field dependence of the magnetization versus flux \ along the path (A-B) is
given on Fig 5a, on which one sees the jump of the magnetic moment to a
positive value when the first order line is crossed. This behavior should be
compared with the smooth one of the magnetization along path (C-B), in the
paramagnetic phase at fixed field and versus temperature, on Fig 5b.

We now give on Fig 6 the magnetization density at points A and B versus
$\rho/R$ , i.e. when the ring is crossed from the inner circle to the outer
one. As already explained in the prevoius section, one clearly sees that the
inner part is paramagnetic while the outer one is diamagnetic and that the
balance of the two results in a small PME effect. Let us stress that,
following this line of argument, and as said for disks in the previous
section, one should observe for disks with very small holes a small domain
with a positive local magnetic moment (around the hole) while there is no
global PME effect.

We discuss now the order parameter distribution across the ring. On Fig 7,
comparing these densities, we find that the order parameter is larger in the
inner part than in the outer part , at point A i.e. in the diamagnetic region.
On the contrary, in the paramagnetic region, at points B , the order parameter
is 30\% larger in the vicinity of the outer edge than around the hole.

Figure 8 shows how the order parameter density profile changes when crossing
the first order transition at $\phi/\phi_{0}\,=1.4$ at fixed temperature
$R^{2}/\xi^{2}=1.2$. We have plotted the values of $|\psi_{J}|$ on the two,
inner and outer, boundaries of the ring for $J=0$ before the transition and
for \ $J=1$ after.

It is worthwhile to notice that the regions of the ring where the order
parameter is the largest, changes as the first order line is crossed, passing
from the inner part before the transition to the outer one after. We have
checked on the example of $d\,/\,R=0.5$ that this result should be general. In
this last case it is observed that, sweeping the temperature range of $J=1$
from one transition point $(J=0\rightarrow J=1)$ to the next one
$(J=1\rightarrow J=2)$, the order parameter profile $|\psi_{1}|$ evolves from
more populated on the inner edge to more populated on the outer one. Let us
note that, even though the superconducting parameter is maximal in this last
region, a paramagnetic effect arises because of the large inner part of the
disk in which the magnetization density is positive (see Fig. 6).

\section{Conclusions}

\bigskip

The paramagnetism of a mesoscopic object can obviously be due to any mechanism
leading to non equilibrium configurations. However, at equilibrium, the PME
depends strongly on the geometry of the sample and presumably can be observed
in many cases much more involded than the toy model presented here. Here we
have shown that the giant vortex state of a pierced disk can exhibit PME in
minimal energy configurations. The size of the hole controls both the number
and extension of the paramagnetic domains. Increase of the field at fixed
temperaure to enter those domains causes the order parameter to jump from
inner to outer localization. The present approach does not explain the PME
seen on the surface superconducting states of 1:4 elongated ellipses
\cite{SupraSurfEllipses} for which the PME in an asymmetric ring\ should be
studied. Paramagnetic Meissner effect in a multiply-connected array of
Josephson junctions has been reported \cite{NCBWL} for which the diamagnetic
current flows on the exterior plaquettes wheras the paramagnetic current flows
in the inside of the sample; this is very similar to that we have described here.

\section{Acknowledgements}

A.\ Buzdin is warmly thanked for his interest in this work and for numerous
and helpfull discussions.

\begin{center}
{\Huge Figure captions }
\end{center}

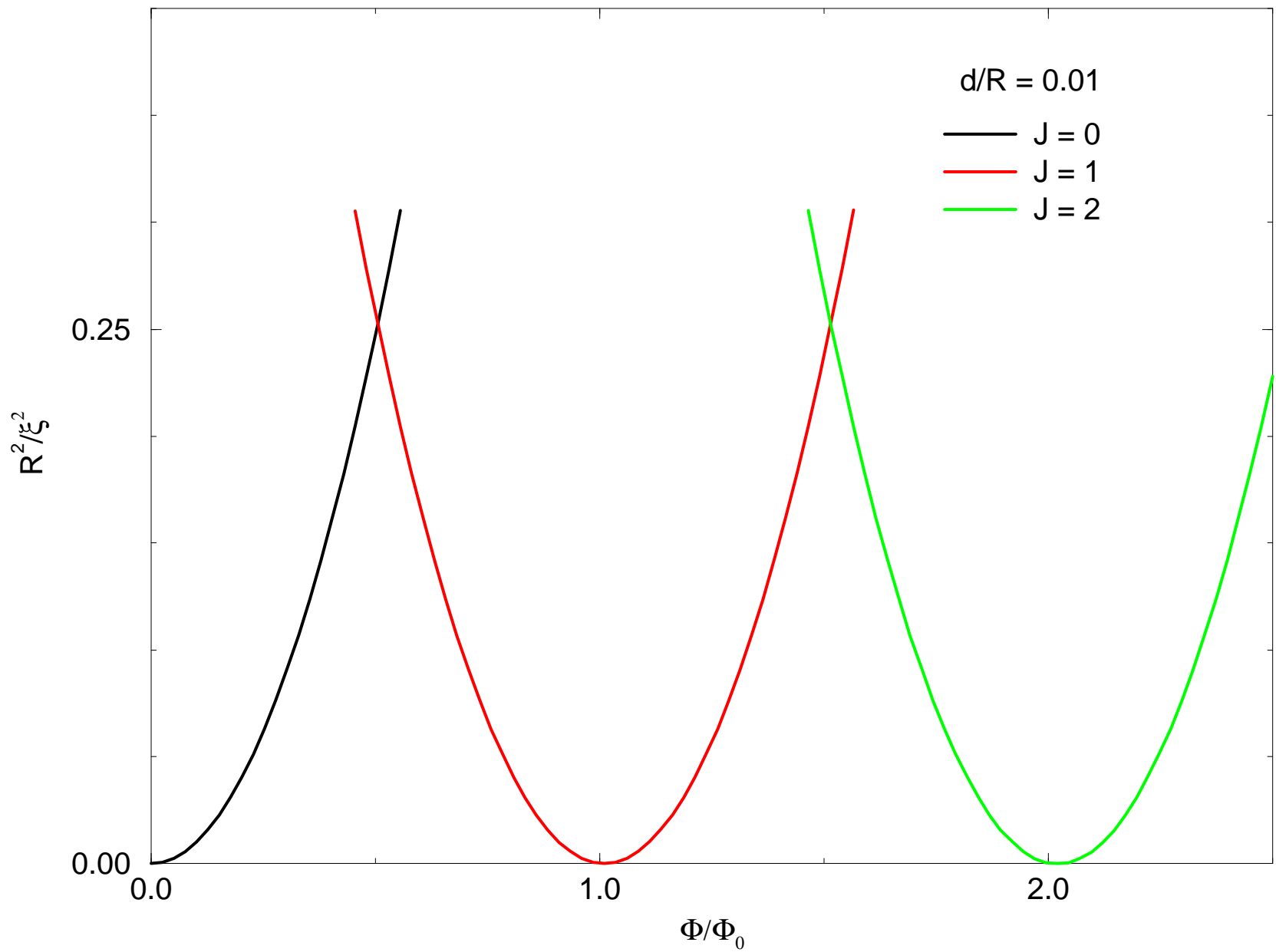
\textbf{Figure 1}. The $\ $phase diagram of a very thin ring with $d/R=0.01$,
\ versus normalized temperaure $R^{2}\,/\,\xi^{2}\thicksim\,\,1-T\,/\,T_{c}%
\,$\ and flux $\phi\,/\,\phi_{0}$ in flux quantum units. The surface
superconducting critical line $Hc_{3\text{ }}$ has been drawn for the first
three angular momentum values. \ \ \ \ \ \ \ \ \ \ \ \ \ \ \ 

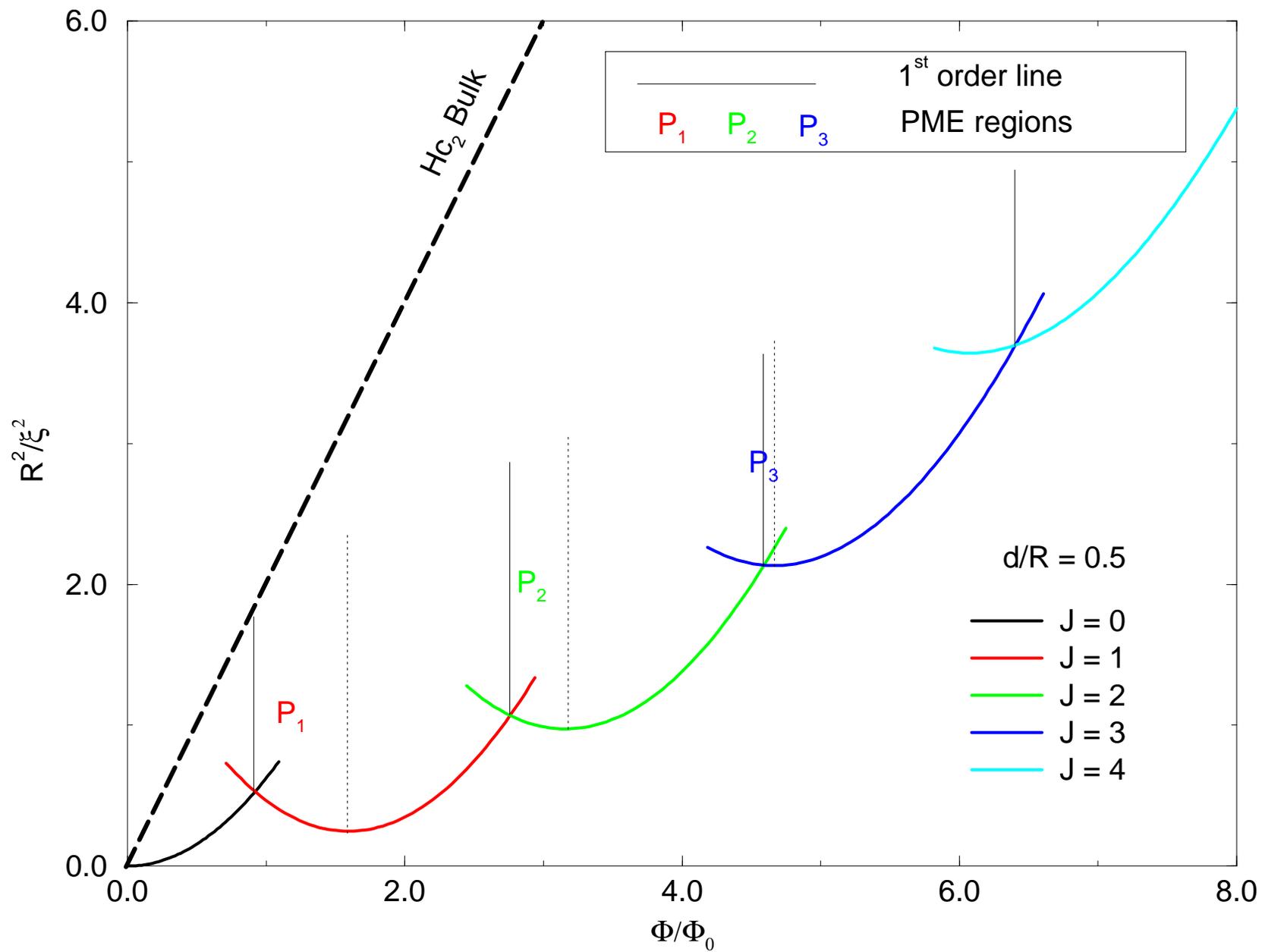
\textbf{Figure 2.} The upper critical field $Hc_{3\text{ }}$for a ring with
$d/R=0.5$. The first order lines which separate domains of different angular
momentum are drawn as thin full vertical lines, whereas vertical dashed lines
mark the expected boundary of each paramagnetic region. The present study is
valid to the right hand side of the $Hc_{2\text{ }}$ line which has been shown
for comparison ( for the bulk material) .

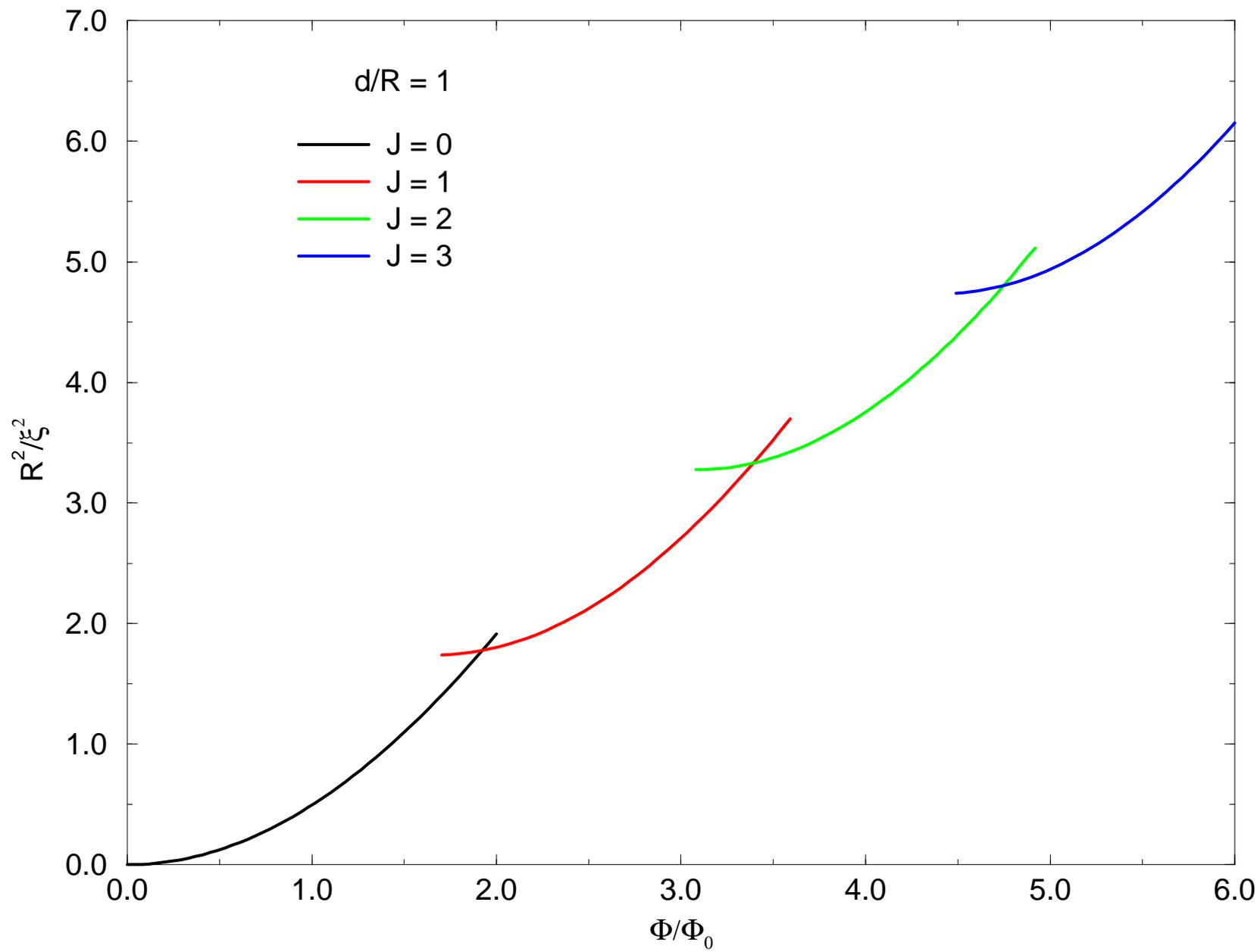
\textbf{Figure 3.} The upper critical field $Hc_{3\text{ }}$for full disk with
$d/R=1$.

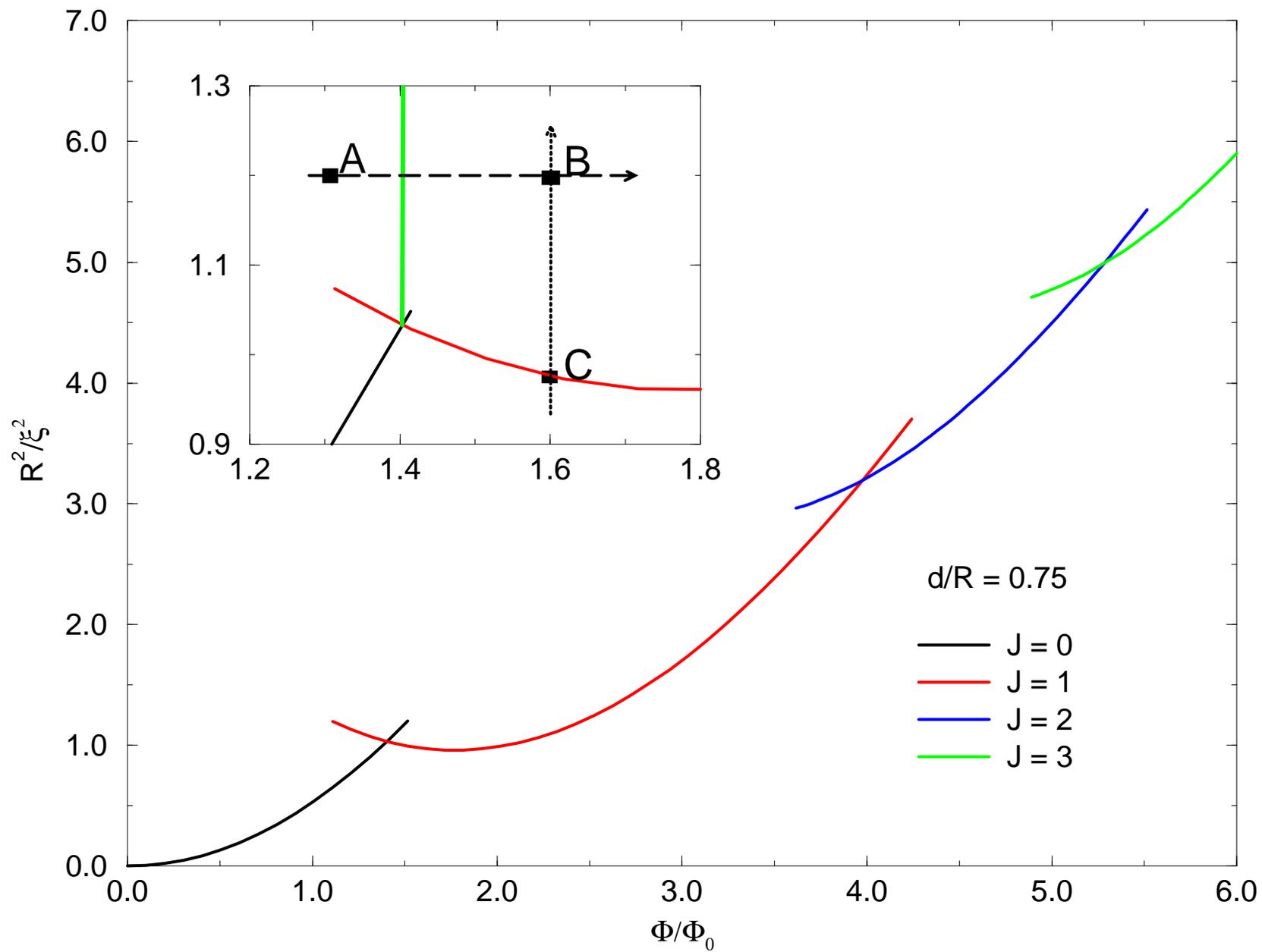
\textbf{Figure 4.} The phase diagram of a disk with a small hole $d/R=0.75$,
in the vicinity of the upper critical fileld $Hc_{3\text{ }}$. The inset plot
shows a zoom of the region $1.2\lesssim\phi/\phi_{0}\,\lesssim\,1.8$ in which
a paramagnetic effect is expected.

\textbf{Figure 5a.} The magnetization at fixed normalized temperature
$R^{2}\,/\,\xi^{2}=1.2$ versus flux along the path \ A-B ( See the inset graph
of Fig. 4 ). The magnetization jumps to small positive values after the transition.

\textbf{Figure 5b. }The magnetization in the paramagnetic region at fixed flux
$\phi/\phi_{0}\,\lesssim\,1.6$ versus temperature, along the path \ C-B ( See
the inset graph of Fig. 4 ).

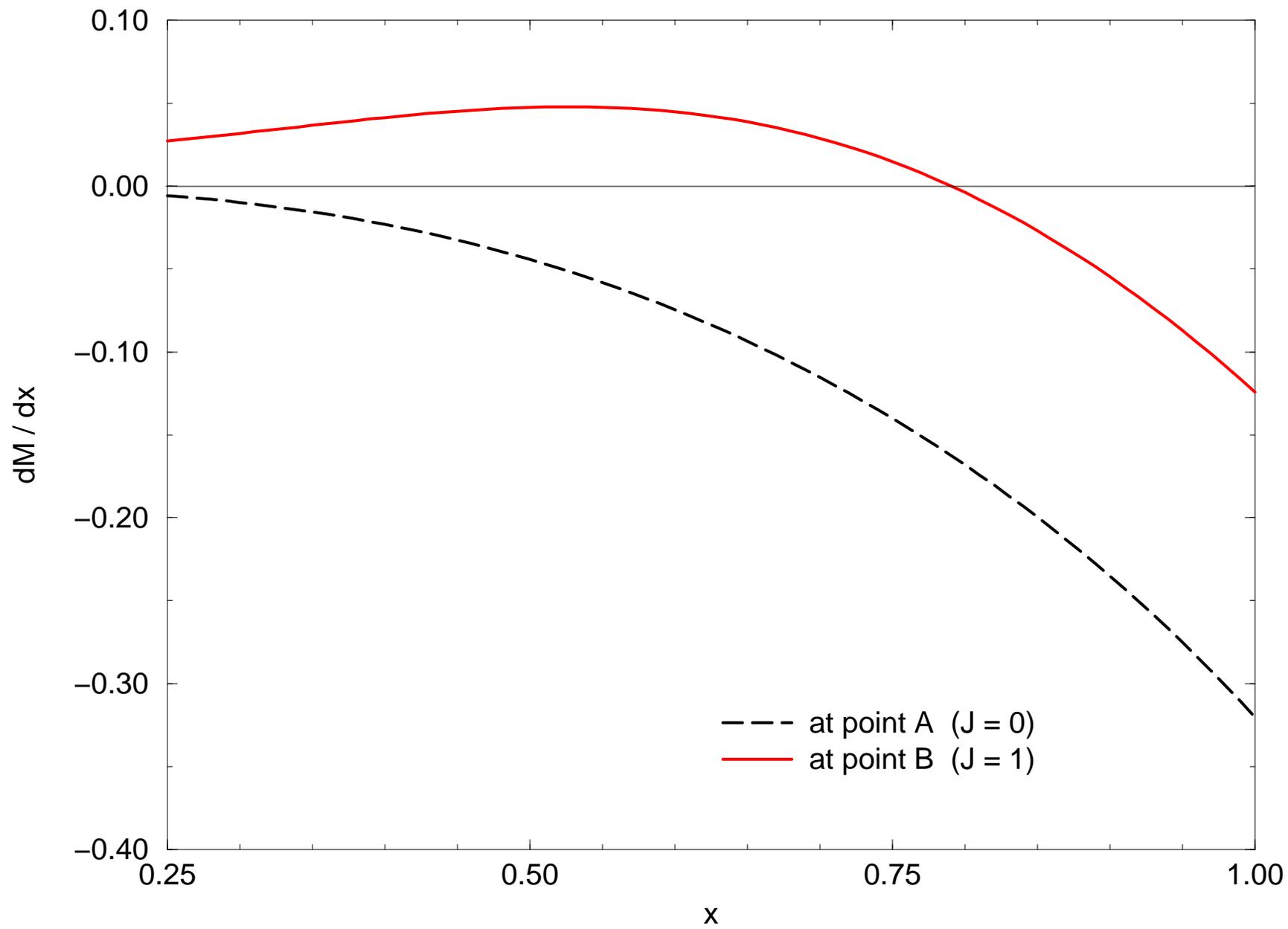
\textbf{Figure 6. \ }The magnetization distribution across the ring, at point
A (full line) and at point B (dashed line) ( for A and B see the inset graph
of Fig. 4 ).

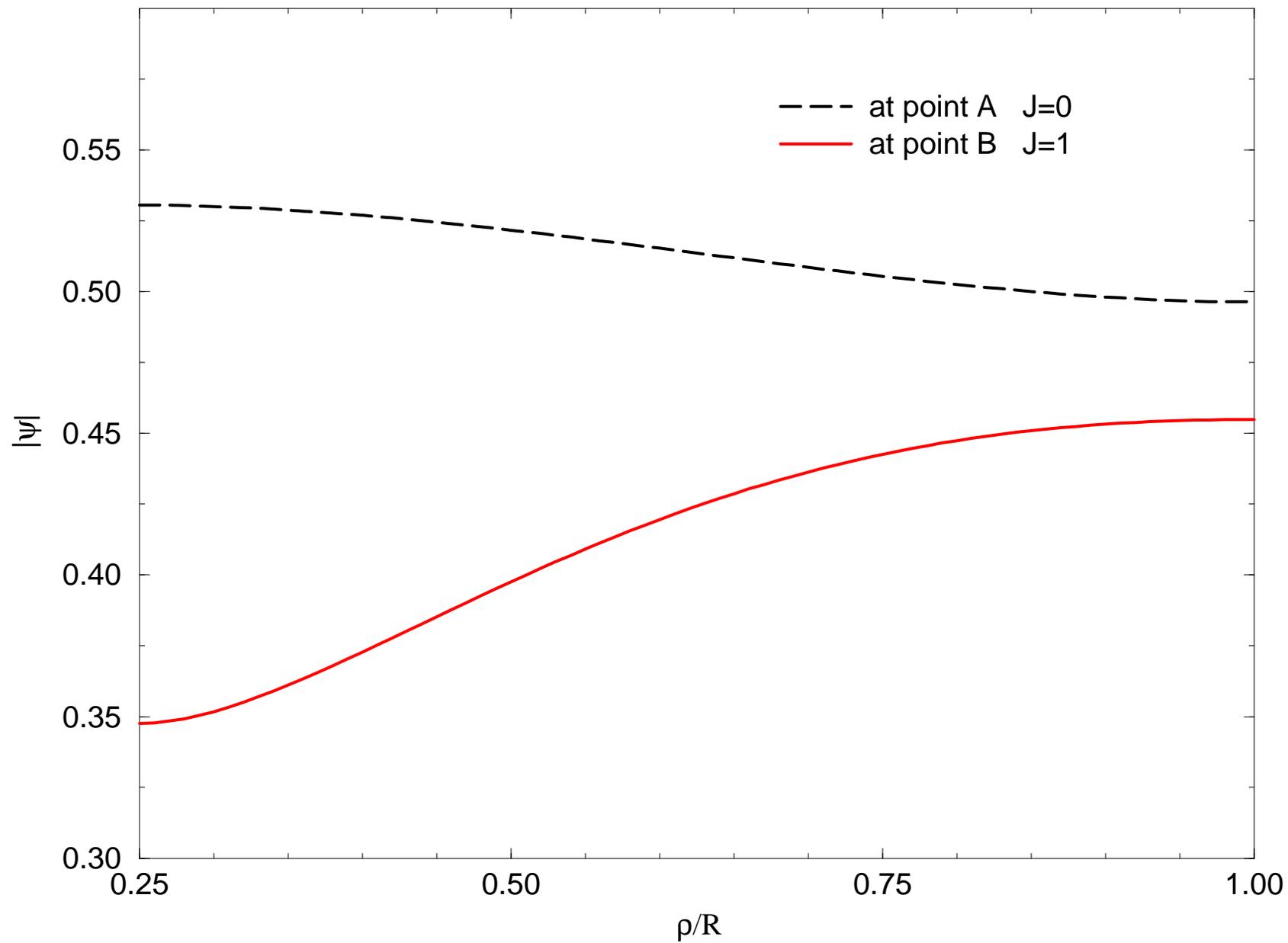
\textbf{Figure 7. \ }Order parameter density at point A, before the transition
, and at point B, in the paramagnetic region, versus the radial variable
$\ 1-d/R<\rho/R<1$ .

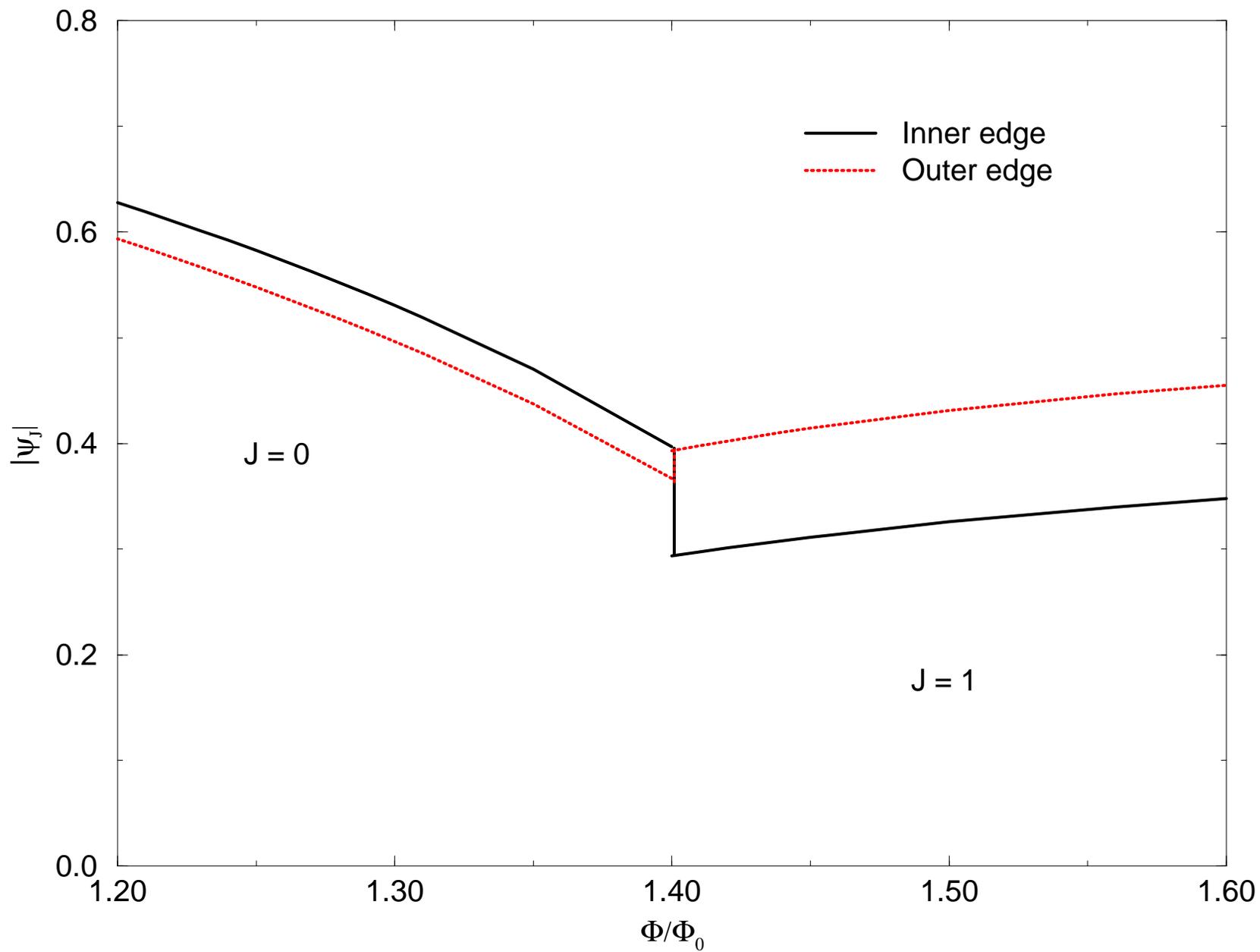
\textbf{Figure 8. }The order parameter density profile oscillation when
crossing the first order transition at $\phi/\phi_{0}=1.4$ for a ring with
$d/R=0.75$. The inner edge is more superconducting than the outer one before
the transition, whereas the reverse situation occurs in the paramagnetic phase.

\begin{thebibliography}{99}
\bibitem{ExpPara}\ D.J. Thompson, M.S.M. Minhaj, L.E. Wenger and J.T. Chen,
Phys. Lett. \textbf{75}, 529 (1995);\newline P. Kostic, A.P. Paulikas, U.
Welp, V.R. Todt, C. Gu, U. Geiser, J.M. Williams, K.D. Carlson and R.A. Klemm,
Phys. Rev. \textbf{B 53}, 791 (1996);\newline A. Terentiev, D.B.\ Watkins and
L.E.D. Long, Phys. Rev. \textbf{B 60}, R761 (1999).

\bibitem{Geim}A. S. Geim, S. V. Dubonos, J. G. S. Lok, M. Henini and J. C.
Mann, Nature (London) \textbf{396}, 144 (1998).

\bibitem{ExpParaHTc}W. Braunisch, N. Knauf, V. Kataev, A. Grutz, A. Kock, B.
Roden, D. Khomskii and D.Wohlleben, Phys. Rev. Lett. \textbf{68}, 1908 (1992)
and\ W. Braunisch, N. Knauf, B. Bauer, A. Kock, A.Becker, B. Freitag, A.
Grutz, V. Kataev, S. Neuhausen , B. Roden, D. Khomskii and D.Wohlleben, Phys.
Rev. \textbf{B 48}, 4030 (1993);\newline B. Schlipe, M. Stindtmann, I.
Nikolic, and K. Baberschke, Phys. Rev.\textbf{\ B 47}, 8331 (1993);\newline S.
Riedling, G. Brauchle, R. Lucht, K. Rohberg, and H. V. Lohneysen, Phys. Rev.
\textbf{B 4}9, 13283 (1994);\newline U. Onbasli, Y.T.\ Wang, A. Naziripour, R.
Tello, W. Kielh, and A.\ M. Hermann, Phys. Stat. Sol. \textbf{B 194}, 371
(1996);\newline G.S. Okram, D.T. Adroja, B.D. Paladia, O. Prakash, and P.A.J.
de Groot, J. Phys.: Cond. Mat. \textbf{9}, L525 (1997).

\bibitem{Ralph}J. von Delft and D.C. Ralph, Phys. Rep. \textbf{345} (2001) 61
;\newline \ \ \ N. Canossa and R. Rossignoli, Physica \textbf{B 320 } (2002) 319.

\bibitem{GrainsP}A.M.\ Clogston, Phys. Rev. Lett. \textbf{9} (1962) 266
;\newline \ \ \ \ \ \ \ B.S.\ Chandrasekhar, Appl. Phys. Lett. \textbf{1}
(1962) 7;\newline \ \ \ \ \ \ \ K. Maki and T. Tsuneto, Prog. Theor. Phys.
\textbf{31} (1964) 945.

\bibitem{metapapers}A.\ E.\ Koshelev and A.\ I.\ Larkin, Phys. Rev. \textbf{B
52} (1995) 13559.\newline V. V. Moschalkov, X. G.\ Qiu and V. Bruyndoncx,
Phys. Rev. \textbf{B 55} (1999) 11793.\newline P.\ S. Deo, V.\ A. Schweigert,
F.\ M.\ Peeters, and A.\ K.\ Geim, Phys. Rev. Lett. \textbf{79} (1997)
4653.\newline P.\ S. Deo, V.\ A. Schweigert and F.\ M.\ Peeters, Phys. Rev.
\textbf{B 59} (1999) 6039.

\bibitem{Zarkov}G.\ F.\ Zarkov, Phys. Rev. \textbf{B 63} (2001) 214502.

\bibitem{LitParks}W. A. Little and R. D. Parks, Phys. Rev. Lett. \textbf{9}
(1962) 9; \ \ Phys. Rev. \textbf{A 133} (1964) 97.

\bibitem{DauMeyBuz}M. Daumens, C. Meyers and A. Buzdin, Phys. Lett.\ \textbf{A
248} (1998) 445.

\bibitem{MeyDau}C. Meyers and M. Daumens, Phys. Rev. \textbf{B 62} (2000) 9762.

\bibitem{BBP}A.Bezryadin, A. Buzdin and B. Pannetier, Phys. Rev. \textbf{B 51}
(1995) 3718.

\bibitem{BGRW}O. Buisson, P. Gandit, R. Rammal, Y. Y.\ Wang and B. Pannetier,
Phys.\thinspace Lett. \textbf{A 150} (1990)\textbf{\ \ }36.

\bibitem{SJdG}D. Saint James and \ P. G. de Gennes, Phys. Lett.\textbf{\ 7}
(1963) 306.

\bibitem{Abra}Handbok of Mathematical Functions, edited by M. Abramowitz and
I. A. Stegun (Dover, New York, 1970), p 504.

\bibitem{SupraSurfEllipses}C. Meyers, M. Daumens and A. Buzdin, Physica
\textbf{C 325 }(1999) 118 \ [\thinspace See Fig 7 for $b/a=0.25$].

\bibitem{NCBWL}A.P.\ Nielsen, A. B. Cawthorne, P. Barbara, F. C. Wellstood, C.
J. Lobb, R. S. Newrock and M. G. Forrester, Phys. Rev. \textbf{B 62} (2000)
1438 .

\bibitem{DeGe}P. G. De Gennes, \textit{Superconductivity of Metals and Alloys}
(Benjamin, New York,1996).\newline 

\bibitem{Abrik}A. A.\ Abrikosov, \textit{Fundamentals of Theory of Metals,}
(North-Holland, Amsterdam, 1988).\newpage\bigskip
\end{thebibliography}
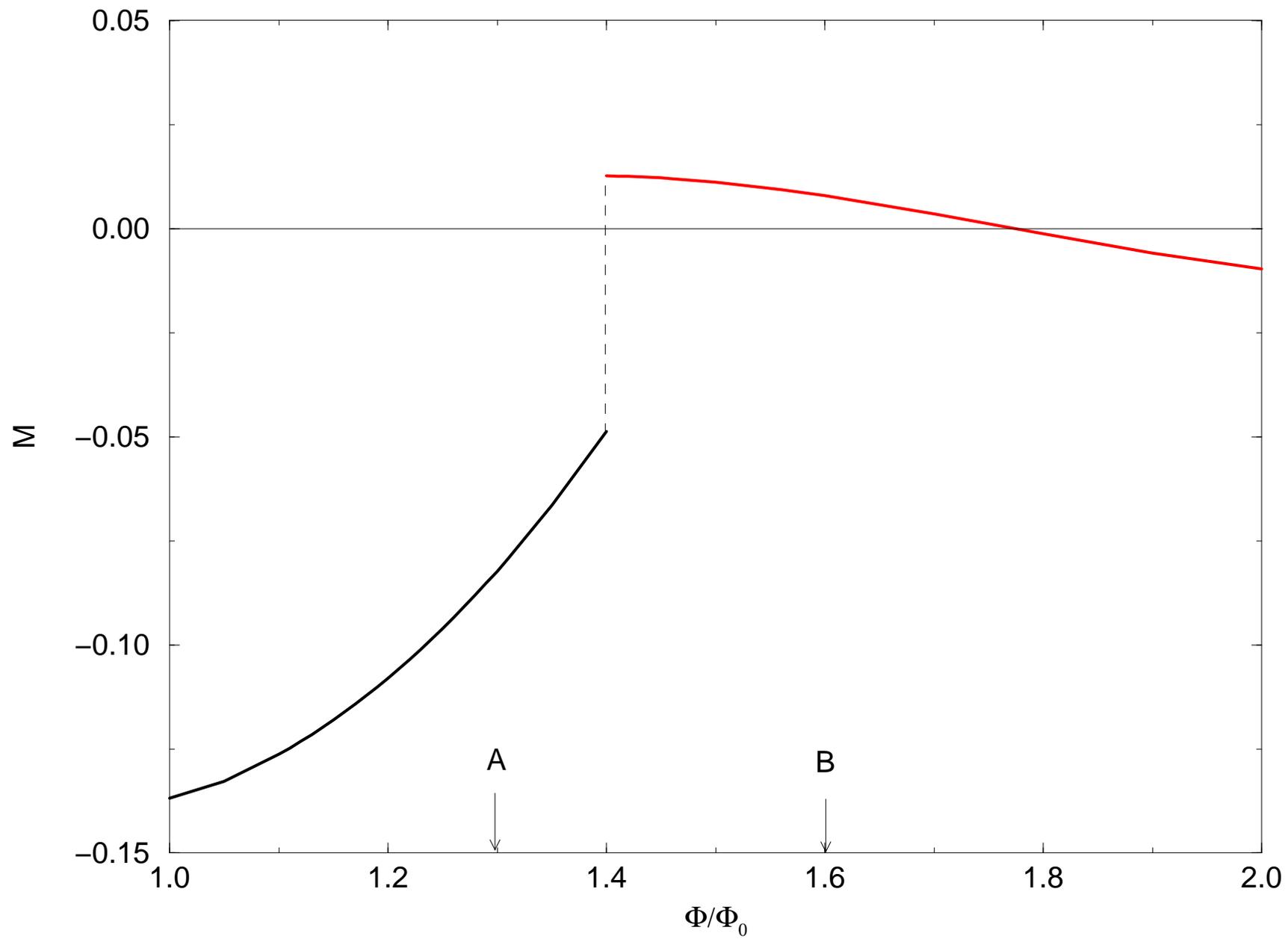
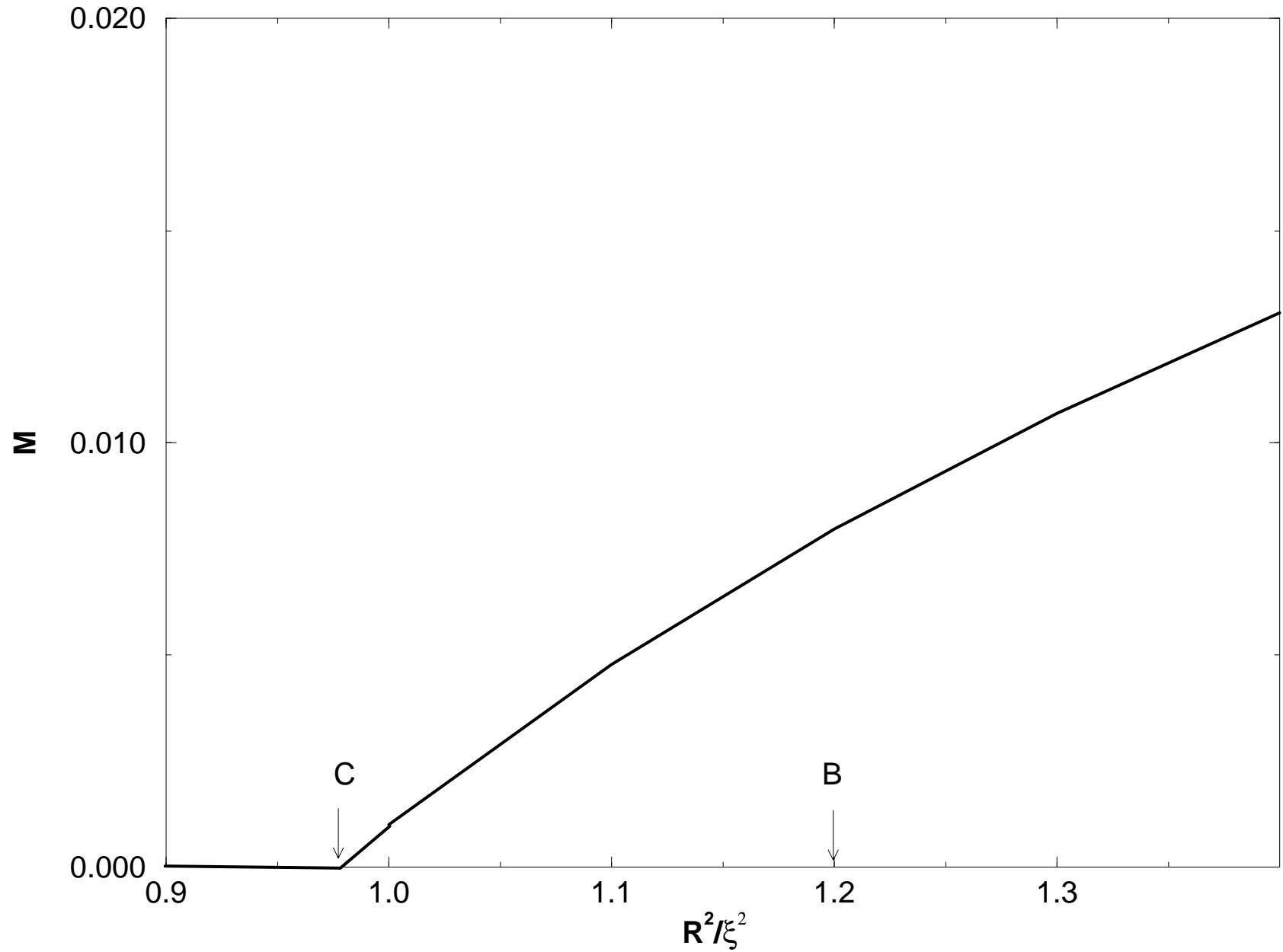
\end{document}